\documentclass[12pt,a4paper,reqno,british]{article}
\usepackage{amsmath,amsfonts,amsthm}

\DeclareTextCompositeCommand{\v}{OT1}{t}{%
    t\kern-.23em\raise.24ex\hbox{'}}

\makeatletter

\providecommand{\LyX}{L\kern-.1667em\lower.25em\hbox{Y}\kern-.125emX\@}

\theoremstyle{plain}    
\newtheorem{thm}{Theorem} 
\theoremstyle{plain}    
\newtheorem{cor}[thm]{Corollary} 
\theoremstyle{plain}    
\newtheorem{lem}[thm]{Lemma} 
\theoremstyle{plain}    
\newtheorem{prop}[thm]{Proposition}
\theoremstyle{remark}
\newtheorem*{rem*}{Remark}
\theoremstyle{remark}
\newtheorem*{rems*}{Remarks}

\theoremstyle{remark}
\newtheorem*{not*}{Notation}

\newcommand{\C}{\mathbb{C}}
\newcommand{\N}{\mathbb{N}}
\newcommand{\R}{\mathbb{R}}
\newcommand{\Z}{\mathbb{Z}}

\makeatother

\begin{document}

\title{Asymptotic properties of the differential equation
  $h^3(h''+h')=1$}

\author{
  J. Asch$^{1}$, R.D. Benguria$^{2}$, P. \v{S}\v{t}ov\'\i\v{c}ek$^3$\\
  \mbox{}\\
  $^1$Centre de Physique Th\'eorique, CNRS, Luminy, Case 907, \\
  Marseille Cedex 9, France\\ asch@cpt.univ-tln.fr; Fax: \# 33 4
91269553\\
  \mbox{}\\
  $^2$ Facultad de F\'{\i}sica, P. U. Cat\'{o}lica de Chile, Casilla
  306, \\
  Santiago 22, Chile \\
  \mbox{}\\
  $^3$Department of Mathematics, Faculty of Nuclear Science, \\
  Czech Technical University, Trojanova 13, 120 00 Prague, \\
  Czech Republic }

\date{1.10. 2001}

\maketitle

\begin{abstract}
  \noindent We derive the form of the asymptotic series, as $t\to +\infty$,
  for a general solution $h(t)$ of the non-linear differential equation
  $h(t)^{3}(h''(t)+h'(t))=1$.
\end{abstract}

\section{Introduction}

The purpose of this article is to describe the asymptotics of
solutions of the second-order ordinary non-linear differential
equation
\begin{equation}
  \label{eq:diff_eq_for_h}
  h(t)^{3}(h''(t)+h'(t))=1
\end{equation}
with initial conditions
\begin{equation}
  \label{eq:init_conds_for_h}
  h(t_0)=h_{0}>0,\textrm{ }h'(t_0)=h_{1}.
\end{equation}
Before formulating the result let us describe our motivation and the
origin of the problem which has its roots in the physical Hall effect.

In a classical mechanics description the issue is to study the
dynamics of a point mass moving in a periodic planar potential and
driven by an exterior electromagnetic field where the magnetic field
is constant and the electric field circular and created by a linearly
time dependent flux tube through the origin, see \cite{Halperin} for
the origin of the model. The equations of motions are Hamiltonian. The
time dependent Hamiltonian is 
\[
H(t,q,p)=\frac{1}{2}\left(p-A(q,t)\right)^2+V(q,t) \hbox{ \rm on }
\R^2\setminus 0\times \R^2
\]
with
\[
A(q,t)=\left(\frac{b}{2}-\frac{et}{\vert
    q\vert^2}\right)(q_2,-q_1).
\]
Here $b$ and $e$ are real parameters and $V$ a smooth periodic function.
In Newtonian form the equations of motion are
\[
\ddot q=E(q)+b\ {\mathbb D }\dot q-\nabla V(q) \quad\hbox{\rm in }
\R^2\setminus \{0\}
\] 
where $\mathbb D$ is rotation by $\pi/2$ and 
$E(q)=|q|^{-2}\,{\mathbb D}q=-\partial_tA(q).$

We shall prove elsewhere that if $b$ and $e$ are nonzero the solutions
are diffusive with or without direction depending on the direction of
the fields. In this article we discuss the particular case when 
$e=1,b=0$ and $V=0$. In polar coordinates the Hamiltonian reads
\[
H(t,r,\phi,p_r,p_\phi)=\frac{1}{2}\left(p_r^2+\frac{1}{r^2}(p_\phi+t)^2
\right)
\] 
and the equations of motion become
\begin{displaymath}
  p_{\phi }'=0,\textrm{ }\phi '=\frac{p_{\phi }+t}{r^{2}},
  \textrm{  }p_{r}'=\frac{(p_{\phi }+t)^{2}}{r^{3}},
  \textrm{ }r'=p_{r}\, .
\end{displaymath}
Consequently, $p_{\phi }$ is a constant and $r''=r^{-3}(p_{\phi
  }+t)^{2}$. After a shift in time we arrive at the equation
\begin{displaymath}
  r''=\frac{t^{2}}{r^{3}}\, .
\end{displaymath} 
The substitution
\begin{displaymath} 
  r(t)=t\, h(\ln t)
\end{displaymath}
leads to equation (\ref{eq:diff_eq_for_h}).

In order to formulate our result we have to introduce some auxiliary
notation.  Let $s_{m,k}\in\R[a_{1},a_{2},\ldots a_{k}]$ be polynomials
defined as follows:
\begin{displaymath}
  s_{m,k}(a_{1},a_{2},\ldots ,a_{k})=\sum _{i_{1}+i_{2}+\ldots+i_{m}=k}
  a_{i_{1}}a_{i_{2}}\ldots a_{i_{m}},
\end{displaymath} 
$m=1,2,\ldots$, $k=0,1,2,\ldots$. Clearly, $s_{m,k}(a_{1},a_{2},\ldots
,a_{k})=0$ if $k<m$, and we set by definition $s_{0,k}=\delta _{0,k}$.
The polynomials obey the recursive rule
\begin{displaymath} 
  s_{m+1,k}(a_{1},a_{2},\ldots
  ,a_{k})=\sum ^{k-1}_{j=m}s_{m,j}(a_{1},a_{2},\ldots ,a_{j})\,
  a_{k-j}\quad \textrm{for }m+1\leq k.
\end{displaymath} 
In the space of formal power series, $\R [[x]]$, it holds
\begin{displaymath} 
  \left( \sum _{k=1}^{\infty }a_{k}x^{k}\right) ^{m}=
  \sum _{k=m}^{\infty }s_{m,k}(a_{1},a_{2},\ldots ,a_{k})\, 
  x^{k},\quad m=0,1,2,\ldots .
\end{displaymath}
This implies that if
\begin{displaymath} 
  a=\sum _{k=1}^{\infty }a_{k}\,x^{k},f=
  \sum_{k=0}^{\infty }f_{k}\,x^{k}\in \R [[x]]
\end{displaymath} 
then
\begin{equation}
  \label{eq:compose_power_ser}
  \sum _{m=0}^{\infty }f_{m}\,a^{m}
  =\sum _{k=0}^{\infty }g_{k}\,x^{k}\quad 
  \textrm{where }g_{k}=\sum _{m=0}^{k}f_{m}s_{m,k}(a_{1},a_{2},
  \ldots ,a_{k})\, .
\end{equation}

Furthermore, set
\begin{equation}
  \label{eq:def-sigma0}
  \sigma ^{0}_{k}(a_{1},a_{2},\ldots ,a_{k})=
  \sum ^{k}_{j=1}\frac{(-1)^{j+1}}{j}\, s_{j,k}(a_{1},a_{2},\ldots ,a_{k})
\end{equation}
and
\begin{equation}
  \label{eq:def-sigma_m}
  \sigma ^{m}_{k}(a_{1},a_{2},\ldots ,a_{k})=
  \sum ^{k}_{j=0}\binom{-m}{j}\, s_{j,k}(a_{1},a_{2},\ldots ,a_{k})\quad
  \textrm{for }m\geq 1.
\end{equation}
Then it holds, in $\R [[x]]$,
\begin{eqnarray*}
  \ln \left( 1+\sum ^{\infty }_{k=1}a_{k}x^{k}\right)  & = & 
  \sum ^{\infty }_{k=1}\sigma ^{0}_{k}(a_{1},a_{2},\ldots ,a_{k})\, x^{k},\\
  \left( 1+\sum ^{\infty }_{k=1}a_{k}x^{k}\right) ^{-m} & = & 1+
  \sum^{\infty }_{k=1}\sigma ^{m}_{k}(a_{1},a_{2},\ldots ,a_{k})\,
  x^{k}\quad \textrm{for }m\geq 1.
\end{eqnarray*}

Let $\{\beta _{n}\}_{n=0}^{\infty }$ be a sequence of real numbers
defined recursively,
\begin{equation}
  \label{eq:def-seq_beta}
  \begin{aligned}
    \beta_0 &= 1,\\
    \beta_{n+1} &= \left( n-\frac{3}{4}\right) \beta _{n}
    +\sum^{n}_{\substack {j,k=0\cr j+k=n+1}}\beta_{j}\beta_{k}
    -\sum^{n}_{\substack {j,k,\ell =0\cr j+k+\ell =n+1}}\beta _{j}
    \beta_{k}\beta _{\ell }\, .
  \end{aligned}
\end{equation}
Here are several first values:
\begin{displaymath}
  \beta _{1}=-\frac{3}{4},\textrm{ }\beta _{2}=-\frac{21}{16},
  \textrm{  }\beta _{3}=-\frac{165}{32},\textrm{ }
  \beta_{4}=-\frac{7245}{256},\textrm{ }\ldots .
\end{displaymath}

For a fixed constant $c\in \R$ we introduce a sequence of polynomials,
$p_{n}(c;z)\in \R [z]$, $n\in \Z _{+}$, by the recursive rule
\begin{equation}
  \label{eq:def-p0}
  p_{0}(c;z)=3z-c
\end{equation}
and
\begin{eqnarray}
  p_{n} & = & 3\,\sigma^{0}_{n}(p_{0},p_{1},\ldots ,p_{n-1})
  +\sum ^{n-1}_{k=1}\frac{4^{k+1}\beta _{k+1}}{k}\, 
  \sigma^{k}_{n-k}(p_{0},p_{1},\ldots ,p_{n-k-1})\nonumber \\
  & & +\frac{4^{n+1}\beta _{n+1}}{n}\, .\label{eq:def-p_n}
\end{eqnarray}
This can be rewritten with the aid of the polynomials $s_{m,k}$,
\begin{eqnarray*}
  p_{n} & = & 3\sum ^{n}_{j=1}\frac{(-1)^{j+1}}{j}\,
  s_{j,n}(p_{0},p_{1},
  \ldots ,p_{n-1})\\
  & & +\, \sum ^{n-1}_{k=0}\frac{4^{n-k+1}\beta _{n-k+1}}{n-k}
  \sum^{k}_{j=0}\binom{-n+\, k}{j}\, s_{j,k}(p_{0},p_{1},\ldots ,p_{k-1}).
\end{eqnarray*}
For $n\geq 1$, the degree of $p_{n}(c;z)$ is less or equal to $n$ (this
can be easily shown by induction when using the fact that for any
monomial $a_{i_{1}}^{\, s_{1}}a_{i_{2}}^{\, s_{2}}\ldots a_{i_{\ell
    }}^{\, s_{\ell }}$ occurring in $\sigma^{m}_{k}(a_{1},a_{2},\ldots
,a_{k})$ it holds $\sum i_{j}s_{j}=k$). Here are several first
polynomials $p_{n}(z)$,
\begin{eqnarray*}
  p_{1}(c;z) & = & 9\, z-21-3\, c,\\
  p_{2}(c;z) & = & -\frac{27}{2}\, z^{2}+(90+9c)\, z-228-30\, c
  -\frac{3}{2}\, c^{2},\\
  p_{3}(c;z) & = & 27\, z^{3}-\left( \frac{621}{2}+27\, c\right) z^{2}
  +(1638+207\, c+9\, c^{2})\, z\\
 &  & -3540-546\, c-\frac{69}{2}\, c^{2}-c^{3}.
\end{eqnarray*}

Now we are able to formulate the result.

\begin{thm}
  \label{thm:MainTheorem}
  For any initial data 
  $(t_{0},h_{0},h_{1})\in \R \times \, ]0,\infty [\,\times\R$ 
  there exists a~unique solution $h(t)$ to the problem
  (\ref{eq:diff_eq_for_h}), (\ref{eq:init_conds_for_h}) on the real
  line.  Moreover, there exists a constant 
  $c=c(t_0,h_{0},h_{1})\in \R$ such that it holds, for all 
  $n\in \Z _{+}$ and $t\to+\infty$:
  \begin{equation}
    \label{eq:asympt_h(t)}
    h(t)=(4t)^{1/4}\left( 1+\sum ^{n}_{k=1}
      \frac{q_{k}(c;\ln(4t))}{t^{k}}
      +O\Big (\left( \frac{\ln (t)}{t}\right)^{n+1}\Big)\right) 
  \end{equation} 
  where
  \begin{displaymath} 
    q_{k}=\sum ^{k}_{m=1}\frac{1}{4^{k}}\,
    \binom{\frac{1}{4}}{m}s_{m,k}(p_{0},p_{1},\ldots ,p_{k-1})\, .
  \end{displaymath} 
  The degree of $q_{k}(c;z)$ is less than or equal to $k$.
\end{thm}
\begin{rems*}
  (i) Several first polynomials $q_{k}(c;z)$ are
  \begin{eqnarray*}
    q_{1}(c;z) & = & \frac{3}{16}\, z-\frac{1}{16}\, c,\\
    q_{2}(c;z) & = & -\frac{27}{512}\, z^{2}+\left( 
      \frac{9}{64}+\frac{9}{256}\, c\right) z-\frac{21}{64}
    -\frac{3}{64}\, c-\frac{3}{512}\, c^{2},\\
    q_{3}(c;z) & = & \frac{189}{8192}z^{3}-\left( 
      \frac{135}{1024}+\frac{189}{8192}\, c\right) z^{2}
    +\left( \frac{549}{1024}+\frac{45}{512}\, c+\frac{63}{8192}\, 
      c^{2}\right) z\\
    &  & -\frac{57}{64}-\frac{183}{1024}\, c-\frac{15}{1024}\, c^{2}
    -\frac{7}{8192}\, c^{3}.
  \end{eqnarray*}

  (ii) In the final step of the proof, in Subsection 
  \ref{sec:general-init-cond}, we shall show the following invariance property 
  of the asymptotic expansion. Set
  \begin{displaymath}
    A_n(c;t)=
   (4t)^{1/4}\left( 1+\sum ^{n}_{k=1}
   \frac{q_{k}(c;\ln (4t))}{t^{k}} \right),
  \end{displaymath}
  with $n\in\Z_+$ and $c,t\in\R$. Then for all $s\in\R$ it holds true that
  \begin{displaymath}
    A_n(c;t+s)=A_n(c-4s;t)+t^{1/4}\,
    O\Big (\left( \frac{\ln (t)}{t}\right) ^{n+1}\Big ) 
    \quad\mathrm{as}\ t\to+\infty.
  \end{displaymath}
\end{rems*}
  
The remainder of the paper contains all necessary steps to prove
Theorem~\ref{thm:MainTheorem}. We shall proceed as follows.  In
Section \ref{sec:Basic_properties} we show the completeness and
derive the first term of the asymptotic series. In Section
\ref{sec:A_reduced_DE_1st_order} we make use of the fact that the
second order differential equation can be reduced to a first order
differential equation and we investigate the asymptotic properties
of the latter equation. These results are used in Section
\ref{sec:Asympt_of_h(t)} to derive the asymptotic properties of the
original second order differential equation and this way we complete
the proof of Theorem \ref{thm:MainTheorem}. Section
\ref{sec:Add_remark} contains an additional remark on the
asymptotics of the Lambert function.

\section{\label{sec:Basic_properties}Basic properties of the differential
  equation}

The differential equation (\ref{eq:diff_eq_for_h}) is equivalent to
the dynamical system
\begin{equation}
  \label{eq:dyn_system_on_R2}
  (x',y')=\left( y,\frac{1}{x^{3}}-y\right) \quad \textrm{on }
  M=\, \, ]0,\infty [\, \times \R .
\end{equation}

\begin{prop}
  The flow of (\ref{eq:dyn_system_on_R2}) is complete and so for all
  initial data $(t_{0},h_{0},h_{1})\in \R \times \, ]0,\infty [\,
  \times \R$ there exists a unique globally defined positive solution
  $h$ of (\ref{eq:diff_eq_for_h}) with initial conditions
  $h(t_{0})=h_{0}$, $h'(t_{0})=h_{1}$.
\end{prop}
\begin{proof}
  We use the following criterion (c.f. \cite[chap.
  2.1.20]{Abraham-Marsden}): \emph{The flow of a $C^{1}$ vector field
    $\xi$ on a manifold $M$ is complete if there is a proper map $f\in
    C^{1}(M,\R )$ which meets the estimate
    \begin{displaymath}
      \exists A>0,B>0,\, \forall p\in M,\quad |\xi \cdot f(p)|\leq A\,
      |f(p)|+B.
    \end{displaymath} }
  In our case $M=\, \, ]0,\infty [\, \times \R$, 
  $\xi=y\partial_{x}+\left(x^{-3}-y\right)\partial_{y}$ 
  and we choose $f(x,y)=x^{2}+y^{2}+x^{-2}$. With this choice we have
  \begin{displaymath} 
    |\xi \cdot f(x,y)|=|2xy-2y^{2}|\leq
    x^{2}+3y^{2}\leq 3\, f(x,y).
  \end{displaymath} 
  Moreover, for any bounded set $S\subset \R$ the inverse image
  $f^{-1}(S)$ is bounded and separated from the border of the
  half-plane $M$: there exists $\varepsilon >0$ such that
  $f^{-1}(S)\subset [\, \varepsilon ,
  $$\infty \, ]\times \R$.  This implies that $f$ is in fact a proper
  map and the proposition is proven.
\end{proof}
\begin{prop}
  Let $(x(t),y(t))$, with $t\in [\, t_{0},\infty [\,$, be a solution
  of the dynamical system (\ref{eq:dyn_system_on_R2}).  Then there
  exists $T\in [\, t_{0},\infty [\,$ such that
  \begin{displaymath} 
    \forall t,s,\, t\geq s\geq T,\quad \sqrt{2}\, (t+c(s))^{1/4}
    \leq x(t) \leq y(s)+\sqrt{2}\, (t+c(s))^{1/4}
  \end{displaymath} 
  where $c(s)=\frac{1}{4}\, x(s)^{4}-s$.
\end{prop}
\begin{proof}
  Set (in this proof) $g(x,y)=x^{-3}-y$ and
  \begin{displaymath} 
    g_{1}(x,y)=-3\, \frac{y}{x^{4}}-\frac{1}{x^{3}}+y
    =-\left( 1+3\, \frac{y}{x}\right) g(x,y)-3\, \frac{y^{2}}{x}
  \end{displaymath} 
  on $M$. Thus $x'=y$, $y'=g(x,y)$ and
  $\frac{d}{dt}g(x(t),y(t))=g_{1}(x(t),y(t))$.
  
  We shall show first that $y'(t)$ is negative for all sufficiently
  large $t$. Note that
  \begin{displaymath} 
    \forall (x,y)\in M,\quad g(x,y)\leq 0\Longrightarrow g_{1}(x,y)<0.
  \end{displaymath} 
  Thus if $y'(s)\leq 0$ then $y'(t)<0$ for all $t>s$.  Hence it is
  sufficient to show that there is at least one $t$ such that
  $y'(t)\leq 0$. Suppose the contrary. Since it holds
  \begin{displaymath}
    y(t)=y(t_{0})+\frac{t-t_{0}}{x(t_{0})^{3}}-\int ^{t}_{t_{0}}
    \left( 3\, \frac{t-s}{x(s)^{4}}+1\right) y(s)\, ds
  \end{displaymath} 
  there exists $s\geq t_{0}$ such that $y(s)>0$. Then both $x(t)$ and
  $y(t)$ are increasing positive functions on the interval $[\,
  s,\infty [\,$ and, in addition, $x(t)<y(t)^{-1/3}$. So the function
  $g(t)=g(x(t),y(t))$ obeys
  \begin{displaymath} 
    \forall t\geq s,\quad
    g'(t)=-\left( 1+3\, \frac{y(t)}{x(t)}\right) g(t)-3\,
    \frac{y(t)^{2}}{x(t)}<-3\, y(s)^{7/3}<0
  \end{displaymath} 
  which clearly contradicts the assumption $g(t)=y'(t)>0$ for all $t$.
  
  Let now $T\geq t_{0}$ be such that $y'(t)=x(t)^{-3}-y(t)<0$ for all
  $t>T$ and fix $s\geq T$. For any $t>T$ we have
  $(\frac{1}{4}x(t)^{4})'=x(t)^{3}y(t)>1$. Consequently, if $t\geq s$
  then
  \begin{displaymath} 
    x(t)\geq \sqrt{2}\left( t+\frac{1}{4}\, x(s)^{4}-s\right) ^{1/4}.
  \end{displaymath}

  To show the other inequality set, for $t\geq s$,
  $z(t)=x(t)-\sqrt{2}\, (t+c)^{1/4}$ where $c=c(s)$. We find that
  \begin{eqnarray*}
    (e^{t}z')' & = & e^{t}(z'+z'')\, =\, e^{t}\left( \frac{1}{x^{3}}
      -(\sqrt{2}\, (t+c)^{1/4})^{-3}+\frac{3\sqrt{2}}{16}\, 
      (t+c)^{-7/4}\right) \\
    & \leq & \frac{3\sqrt{2}}{16}\, e^{t}\, (t+c)^{-7/4}.
  \end{eqnarray*}
  It follows that
  \begin{displaymath} 
    z'(t)\leq e^{s-t}z'(s)+\frac{3\sqrt{2}}{16}\,
    e^{-t}\int _{s}^{t}e^{u}\, (u+c)^{-7/4}\, du
  \end{displaymath} 
  and
  \begin{eqnarray*}
    z(t) & \leq  & z(s)+\left( 1-e^{s-t}\right) z'(s)
    +\frac{3\sqrt{2}}{16}\, \int _{s}^{t}e^{-\tau }\int _{s}^{\tau
      }e^{u}\, 
    (u+c)^{-7/4}\, du\, d\tau \\
    & \leq  & z(s)+z'(s)+\frac{3\sqrt{2}}{16}\, \int _{s}^{t}
    \left( 1-e^{u-t}\right) \, (u+c)^{-7/4}\, du\\
    & \leq  & z(s)+z'(s)+\frac{3\sqrt{2}}{16}\, \frac{4}{3}\, (s+c)^{-3/4}\\
    & = & z(s)+x'(s).
  \end{eqnarray*}
  But $z(s)=0$ and so $x(s)\leq y(s)+\sqrt{2}\, (t+c)^{1/4}$.
\end{proof}

\begin{cor}\label{cor:hprime_positive}
  If $h(t)$ is a solution of (\ref{eq:diff_eq_for_h}) on $[\,
  t_{0},\infty [\,$ with the initial conditions $h(t_{0})=h_{0}>0$,
  $h'(t_{0})=h_{1}$, then there exists $T\geq t_{0}$ such that
  $h'(t)>0$ for all $t>T$.
\end{cor}

\begin{cor}
  If $h(t)$ is a solution of (\ref{eq:diff_eq_for_h}) on $[\,
  t_{0},\infty [\,$ with the initial conditions $h(t_{0})=h_{0}>0$,
  $h'(t_{0})=h_{1}$, then
  \begin{displaymath} 
    h(t)=\sqrt{2}\, t^{1/4}+O(1)\quad \textrm{as }t\to +\infty .
  \end{displaymath}
\end{cor}

\begin{rem*}
  This means that if we restrict ourselves in what follows to the
  initial condition (\ref{eq:init_conds_for_h}) with $h_{1}>0$ we
  don't loose the generality as far as the asymptotics is concerned.
  Furthermore, owing to the invariance of the differential equation in
  time we can set $t_{0}=0$. This fact will be used in the course of
  the proof. First we verify Theorem \ref{thm:MainTheorem} for the
  particular case when $t_0=0$ and $h_1>0$ and then, in 
  Subsection \ref{sec:general-init-cond}, we shall extend the result
  to the general initial condition.
\end{rem*}

\section{\label{sec:A_reduced_DE_1st_order}A reduced differential equation
  of first order }

In accordance with the remark at the end of 
Section \ref{sec:Basic_properties} we assume that $t_0=0$ and
$h_1>0$. The second-order differential equation equation
(\ref{eq:diff_eq_for_h}) is invariant in $t$ and this is why it can be
reduced to a first-order differential equation. Actually, using the
substitution $h(t)=\bigl(G^{-1}(4t)\bigr)^{1/4}$, 
$z_{0}=4/h_{0}^{\, 4}$ and $g_{0}=h_{0}^{\, 3}h_{1}$ where
\begin{displaymath} 
  G(x)=\int^{x}_{h_{0}^{4}}\frac{ds}{g\left( \frac{4}{s}\right) }
\end{displaymath}
we arrive at a first-order nonlinear differential equation, namely 
\begin{equation}
  \label{eqationA}
  \left( 1-\frac{3}{4}z\, g(z)-z^{2}\, g'(z)\right) g(z)=1,\, \, 
  g(z_{0})=g_{0}.
\end{equation}
We shall carry out the computations relating (\ref{eq:diff_eq_for_h}),
(\ref{eq:init_conds_for_h}) and (\ref{eqationA}) later, in Subsection
\ref{sec: reduction_second_order_DE}. Here we concentrate on the study
of (\ref{eqationA}) on the interval $\lbrack\,0,z_0\,\rbrack$, assuming 
that $z_{0}>0$ and $g_{0}>0$.

\subsection{Domain of the left maximal solution \protect$g(z)\protect$}

We shall need two equivalent forms of the differential equation,
namely
\begin{equation}
  \label{eqationB}
  g'(z)=\frac{1}{z^{2}}\left( 1-\frac{1}{g(z)}\right) 
  -\frac{3}{4}\frac{g(z)}{z}
\end{equation}
and
\begin{equation}
  \label{eqationC}
  \left( z^{3/4}g(z)\right) '=z^{-5/4}\left( 1-\frac{1}{g(z)}\right) \, .
\end{equation}

\begin{rem*}
  In what follows we use repeatedly the following elementary argument:
  \emph{if $\psi$ and $\varphi$ are two differentiable functions on
    $]a,b[$ and the equality $\psi (z)=\varphi (z)$ implies $\psi
    '(z)<\varphi '(z)$, for all $z\in \, ]a,b[$, then the two
    functions coincide in at most one point} $z\in \, ]a,b[$.
\end{rem*}
\begin{prop}
  \label{L:limg}Let $g(z)$ be the left maximal solution of
  (\ref{eqationA}). Then $g(z)$ is defined and positive on $]0,z_{0}\,
  ]$, and
  \begin{equation}
    \label{limg}
    \lim _{z\downarrow 0}g(z)=1\, .
  \end{equation}
\end{prop}
\begin{proof}
  Let $\gamma$ be the minimal non-negative number such that $g(z)$ is
  defined on $]\gamma ,z_{0}\, ]$. Equation (\ref{eqationA}) clearly
  excludes the possibility that $g(z)=0$ for some $z\in \, ]\gamma
  ,z_{0}\, ]$. So $g(z)$ is positive on $]\gamma ,z_{0}\, ]$. Our goal
  is to show that $\gamma =0$ and (\ref{limg}) holds true. We split
  the proof into six claims.
  
  \emph{(i)} $\exists \xi \in \, ]\gamma ,z_{0}\, ]$ s.t.  $g(\xi
  )\geq 1$.
  
  Suppose that $g(z)<1$, $\forall z\in \, ]\gamma ,z_{0}\, ]$.  Then,
  by (\ref{eqationB}), $g'(z)<0$ on $]\gamma ,z_{0}\, ]$ and so there
  exists $\lim _{z\downarrow \gamma }g(z)=g_{1}$ with $g_{0}<g_{1}\leq
  1$. Hence, by minimality of $\gamma$, it should hold $\gamma =0$.
  According to (\ref{eqationC}),
  \begin{displaymath} 
    \left( z^{3/4}g(z)\right) '<0\Longrightarrow g(z)>g_{0}
    \left( \frac{z_{0}}{z}\right) ^{3/4},\quad 
    \forall z\in \, ]0,z_{0}\, ],
  \end{displaymath} 
  a contradiction. In the following choose $\xi\in\,\,]\gamma,z_0\,]$ 
  to be the largest number such $g(\xi)\geq1$.
  
  \emph{(ii)} $g(z)>1$, $\forall z\in \, ]\gamma ,\xi [$.
  
  Actually, $g(z)=1$ implies $g'(z)<0$.
  
  \emph{(iii)} $\gamma =0$.
  
  If $\gamma >0$ then, by (\ref{eqationC}), $|(z^{3/4}g(z))'|\leq
  \gamma ^{-5/4}<\infty$ for all $z\in \, ]\gamma ,\xi [$. This means
  that $z^{3/4}g(z)$ is absolutely continuous on $]\gamma ,\xi [$ and
  so $\lim _{z\downarrow \gamma }g(z)$ exists and is finite, a
  contradiction with the minimality of $\gamma$.
  
  \emph{(iv)} $\exists \eta \in \, ]0,\xi [$ s.t. $g'(z)\neq 0 $,
  $\forall z\in \, ]0,\eta [$.
  
  Equalling the RHS of (\ref{eqationB}) to zero we get a quadratic
  equation with respect to $g(z)$. Its solution is a couple of
  functions,
  \begin{displaymath} 
    \varphi _{1}(z)=\frac{2}{1+\sqrt{1-3z}},\quad 
    \varphi_{2}(z)=\frac{2}{1-\sqrt{1-3z}},
  \end{displaymath} 
  defined on the interval $]0,\xi [\, \cap \, ]0,\frac{1}{3}[$.
  Clearly, $\varphi _{1}'(z)>0$ and $\varphi _{2}'(z)<0$ everywhere on
  that interval. This implies that the RHS of (\ref{eqationB})
  vanishes in a point $z$ from that interval if and only if
  $g(z)=\varphi _{1}(z)$ or $g(z)=\varphi _{2}(z)$ and in such a case
  either $0=g'(z)<\varphi_{1}'(z)$ or $0=g'(z)>\varphi _{2}'(z) $.
  Thus $\varphi_{1}(z)$ coincides with $g(z)$ in at most one point
  $z$, and the same is true for $\varphi _{2}(z)$.  Consequently,
  there exists $\eta \in \, ]0,\xi [\, \cap \, ]0,\frac{1}{3}[$ such
  that $g'(z)$ doesn't vanish on $]0,\eta [$. Choose $\eta$ having
  this property.
  
  \emph{(v)} $g'(z)>0$, $\forall z\in \, ]0,\eta [$.
  
  If $g'(z)<0$, $\forall z\in \, ]0,\eta [$, then 
  $g_{1}=\lim_{z\downarrow 0}g(z)$ exists (finite or infinite) and 
  $g_{1}>g(\eta)>1$. On the other hand, in virtue of (\ref{eqationB}),
  \begin{equation}
    \label{gprimeneg}
    z^{2}g'(z)=1-\frac{1}{g(z)}-\frac{3}{4}\,z\, g(z)<0\, .
  \end{equation}
  Equality (\ref{eqationC}) implies that $(z^{3/4}g(z))'>0$ on
  $]0,\eta [$ and so
  \begin{displaymath} 
    g(z)<g(\eta )\left( \frac{\eta }{z}\right)^{3/4},\quad 
    \forall z\in \, ]0,\eta [\, .
  \end{displaymath} 
  Consequently, $\lim_{z\downarrow 0}z\, g(z)=0$. Sending $z$ to 0 in
  (\ref{gprimeneg}) gives
  \begin{displaymath} 
    1-\frac{1}{g_{1}}\leq 0\, ,
  \end{displaymath} 
  a contradiction.
  
  \emph{(vi)} $\lim _{z\downarrow 0}g(z)=1$.
  
  From Claim (v) follows that $g_{1}=\lim _{z\downarrow 0}g(z)$ exists
  and $1\leq g_{1}<g(\eta )$. Suppose that $g_{1}>1$.  Then one
  concludes from (\ref{eqationB}) that there exists $\delta \in \,
  ]0,\eta [$ s.t.
  \begin{displaymath} 
    g'(z)>\frac{d}{z^{2}},\, \, 
    \forall z\in \, ]0,\delta [\, ,\quad \mathrm{where}\, \, d
    =\frac{1}{2}\left( 1-\frac{1}{g_{1}}\right) >0.
  \end{displaymath} 
  This implies
  \begin{displaymath} 
    g(z)<g(\delta )+\frac{d}{\delta }-\frac{d}{z},\, \, 
    \forall z\in \, ]0,\delta [\, ,
  \end{displaymath} 
  a contradiction.
\end{proof}

\begin{cor}
  The maximal solution $g(z)$ satisfies the integral identity
  \begin{equation}
    \label{intident}
    g(z)^{2}=2z^{-3/2}\int^{z}_{0}s^{-1/2}(g(s)-1)\, ds,\quad 
    \forall z\in \, ]0,z_{0}\, ].
  \end{equation}
\end{cor}
\begin{proof}
  Rewrite (\ref{eqationB}) as
  \begin{displaymath} 
    (z^{3/2}g^{2})'=2z^{-1/2}(g-1),
  \end{displaymath} 
  and integrate from 0 to $z$ when taking into account (\ref{limg}).
\end{proof}

\subsection{Asymptotics of the left maximal solution \protect$g(z)\protect$
  at \protect$z=0\protect$}

\begin{lem}
  \label{L:prepasymptser}If $\beta \geq 0$ then
  \begin{displaymath}
    e^{-\frac{1}{z}}\int ^{1}_{z}s^{\beta -2}e^{\frac{1}{s}}\,
    ds=O(z^{\beta })\quad \text {as}\, \, z\downarrow 0\, .
  \end{displaymath}
\end{lem}
\begin{proof}
  \begin{eqnarray*}
    e^{-\frac{1}{z}}\int ^{1}_{z}s^{\beta -2}e^{\frac{1}{s}}\, ds 
    & = & e^{-\frac{1}{z}}\left( \int ^{\frac{1}{2z}}_{1}
      +\int ^{\frac{1}{z}}_{\frac{1}{2z}}\right) u^{-\beta }e^{u}\, du\\
    & \leq  & e^{-\frac{1}{z}}\left( e^{\frac{1}{2z}}-e\right) 
    +(2z)^{\beta }e^{-\frac{1}{z}}\left( e^{\frac{1}{z}}
      -e^{\frac{1}{2z}}\right) \, .
  \end{eqnarray*}
\end{proof}
\begin{not*} 
  $(\alpha )_{0}=1$, $(\alpha )_{j} =\alpha (\alpha +1)\dots (\alpha
  +j-1)$.
\end{not*}

\begin{lem}
  \label{L:asymptser}
  Let $\delta >0$, $\nu \in \R $. Then the expression
  \begin{displaymath} 
    z^{-\nu }e^{-1/z}\int ^{\delta }_{z}s^{\nu -2}e^{1/s}\, ds\, ,
  \end{displaymath} 
  regarded as a function in the variable $z$, has the asymptotic
  series, as $z\downarrow 0$,
  \begin{displaymath} 
    \sum ^{\infty }_{j=0}(\nu )_{j}z^{j}\, .
  \end{displaymath}
\end{lem}
\begin{proof}
  It suffices to show that it holds, for all $n\in \Z _{+}$,
  \begin{eqnarray*}
    e^{-\frac{1}{z}}\int ^{\delta }_{z}s^{\nu -2}e^{\frac{1}{s}}\, ds 
    & = & -e^{\frac{1}{\delta }-\frac{1}{z}}
    \sum ^{n-1}_{j=0}(\nu )_{j}\delta ^{\nu +j}
    +\sum ^{n-1}_{j=0}(\nu )_{j}z^{\nu +j}\\
    & & +(\nu )_{n}e^{-\frac{1}{z}}\int ^{\delta }_{z}
    s^{\nu +n-2}e^{\frac{1}{s}}\, ds\, .
  \end{eqnarray*}
  Actually, according to Lemma \ref{L:prepasymptser} this means that
  the relation
  \begin{displaymath} 
    z^{-\nu }e^{-\frac{1}{z}}
    \int^{\delta }_{z}s^{\nu -2}e^{\frac{1}{s}}\,ds
    =\sum ^{n-1}_{j=0}(\nu )_{j}z^{j}+O(z^{n})
  \end{displaymath}
  holds true for all $n\geq -\nu$, and so for all $n\in \N$.
  
  We shall proceed by induction in $n$. For $n=0$ this is a trivial
  equality. The induction step $n\to n+1$:
  \begin{eqnarray*}
    e^{-\frac{1}{z}}\int ^{\delta }_{z}s^{\nu +n-2}e^{\frac{1}{s}}\,
    ds 
    & = & -e^{-\frac{1}{z}}\int ^{\delta }_{z}s^{\nu +n}
    \left( e^{\frac{1}{s}}\right) '\, ds\\
    & = & -e^{-\frac{1}{z}}[s^{\nu +n}e^{\frac{1}{s}}]^{\delta }_{z}
    +(\nu +n)e^{-\frac{1}{z}}\int ^{\delta }_{z}s^{\nu +n-1}
    e^{\frac{1}{s}}\, ds\, .
  \end{eqnarray*}
\end{proof}
Let us consider equation (\ref{eqationA}) (without the initial
condition) in the space of formal power series $\C [[z]]$. Its
solution
\begin{equation}
  \label{tildeg}
  \tilde{g}(z)=\sum ^{\infty }_{k=0}\alpha _{k}z^{k}\in \C [[z]]
\end{equation}
is unique, with the coefficients being determined by the recursive
relation
\begin{equation}
  \label{recurcoeff}
  \alpha _{0}=1,\quad \alpha _{k+1}=\left(
    \frac{1}{2}k+\frac{3}{4}\right) 
  \sum _{j=0}^{k}\alpha _{j}\alpha _{k-j}\quad \textrm{for }k\geq 0\, .
\end{equation}
Several first coefficients are
\begin{displaymath} 
  \alpha _{0}=1,\textrm{ }\alpha_{1}=\frac{3}{4},\textrm{ }\alpha _{2}
  =\frac{15}{8},\textrm{ }\alpha_{3}=\frac{483}{64},\dots \, .
\end{displaymath}

\begin{prop}
\label{P:asymptofg}
The left maximal solution $g(z)$ has an asymptotic series, as
$z\downarrow 0$, that is equal to
\begin{displaymath} 
  \sum^{\infty }_{k=0}\alpha _{k}z^{k}\, .
\end{displaymath}
\end{prop}
\begin{proof}
  We have to show that, for all $n\in \Z _{+}$,
  \begin{equation}
    \label{gasymptordn}
    g(z)=\sum ^{n}_{k=0}\alpha _{k}z^{k}+o(z^{n})\, .
  \end{equation}
  We shall proceed by induction in $n$. The case $n=0$ means that
  $\lim _{z\downarrow 0}g(z)=1$ and is covered by Proposition
  \ref{L:limg}. Let us suppose that (\ref{gasymptordn}) is valid for
  some $n\in \Z _{+}$. Denote (in this proof)
  \begin{displaymath} 
    a_{-}=\liminf_{z\downarrow 0}\frac{g(z)-\sum ^{n}_{k=0}
      \alpha _{k}z^{k}}{z^{n+1}}\, .
  \end{displaymath} 
  Similarly, $a_{+}$ designates the limes superior, as $z\downarrow
  0$, of the same function. Thus the induction step $n\to n+1$ means
  to verify that $a_{-}=a_{+}=\alpha _{n+1}$. We shall do it in three
  steps.
  
  \emph{(i)} $a_{-}\leq \alpha _{n+1}\leq a_{+}$.
  
  If $b<a_{-}$ then there exists $\delta >0$ s.t.
  $g(z)>\sum^{n}_{k=0}\alpha _{k}z^{k}+b\, z^{n+1}$, $\forall z\in \,
  ]0,\delta [$. Further, from the recurrent relation
  (\ref{recurcoeff}) one finds that
  \begin{equation}
    \label{tildeg2}
    \tilde{g}(z)^{2}=\sum ^{\infty }_{k=0}\left( \sum ^{k}_{j=0}\alpha_{j}
      \alpha _{k-j}\right) z^{k}
    =\sum^{\infty }_{k=0}\frac{2}{k+\frac{3}{2}}\, 
    \alpha _{k+1}z^{k}\, ,
  \end{equation}
  and the assumption (\ref{gasymptordn}) implies that the asymptotics
  of $g(z)^{2}$ is given by a truncation of the formal power series
  $\tilde{g}(z)^{2}$, namely
  \begin{displaymath} 
    g(z)^{2}=\sum^{n}_{k=0}\frac{2}{k+\frac{3}{2}}
    \alpha_{k+1}z^{k}+o(z^{n})\, .
  \end{displaymath}
  Combining this with (\ref{intident}) leads to the conclusion that,
  $\forall z\in \, ]0,\delta [$,
  \begin{eqnarray*}
    \sum ^{n}_{k=0}\frac{2}{k+\frac{3}{2}}\, \alpha _{k+1}z^{k}+o(z^{n})
    & > & 2\, z^{-3/2}\int ^{z}_{0}s^{-1/2}
    \left( \sum ^{n}_{k=1}\alpha _{k}s^{k}+b\, s^{n+1}\right) \, ds\\
    & = & 2\sum ^{n}_{k=1}\frac{\alpha _{k}}{k+\frac{1}{2}}\,
    z^{k-1}+2\, \frac{b}{n+\frac{3}{2}}\, z^{n}\, .
  \end{eqnarray*}
  Hence $b\leq \alpha _{n+1}$ for all $b<a_{-}$ and consequently
  $a_{-}\leq \alpha _{n+1}$. The inequality $a_{+}\geq \alpha_{n+1}$
  can be proven symmetrically.
  
  \emph{(ii)} Let $\lambda >0$ be a fixed parameter. In the second
  step we introduce an auxiliary function $\varphi (z)$ whose choice
  depends on whether $n=0$ or $n>0$. In the former case we set
  \begin{eqnarray}
    \varphi (z) & = & \frac{1}{\lambda }z^{-\frac{3}{4}}
    e^{-\frac{1}{\lambda z}}\int ^{1}_{z}s^{-\frac{5}{4}}
    e^{\frac{1}{\lambda s}}\, ds\nonumber \\
    & = & (\lambda z)^{-\frac{3}{4}}e^{-\frac{1}{\lambda z}}
    \int^{\lambda }_{\lambda z}s^{-\frac{5}{4}}e^{\frac{1}{s}}\, ds
    \label{phiint0}
  \end{eqnarray}
  and in the latter one
  \begin{equation}
    \label{phiintn}
    \varphi (z)=z^{-\frac{3}{4}}e^{-\frac{1}{z}}\int ^{1}_{z}
    \sum^{n+1}_{k=0}\varphi_{k}\, s^{\frac{3}{4}+k-2}\, e^{\frac{1}{s}}\, ds
  \end{equation}
  where
  \begin{equation}
    \label{coeffphik}
    \begin{aligned}
      \varphi_0 &= \alpha_0, \\
      \varphi_k &= \alpha_k-\left(k-\frac14\right)\alpha_{k-1}\quad
      \text{for } 1\leq k\leq n,\\
      \varphi_{n+1} &= \lambda\,\alpha_{n+1}-
      \left(n+\frac34\right)\alpha_n.
    \end{aligned}
  \end{equation}
  
  Observe that $\varphi _{1}=\alpha _{1}-\frac{3}{4}\alpha _{0}=0$.
  In the case $n=0$, $\varphi (z)$ solves the differential equation
  \begin{equation}
    \label{phidiffeq0}
    \varphi '(z)=\frac{1}{\lambda \, z^{2}}(\varphi (z)-1)-\frac{3}{4}
    \frac{\varphi (z)}{z}\, ,
  \end{equation}
  while in the case $n>0$, $\varphi (z)$ solves
  \begin{equation}
    \label{phidiffeqn}
    \varphi '(z)=\frac{1}{z^{2}}(\varphi (z)-1)-\frac{3}{4}
    \frac{\varphi (z)}{z}-\sum ^{n+1}_{k=2}\varphi _{k}z^{k-2}\, .
  \end{equation}
  We claim that, in the both cases, the asymptotic behaviour of
  $\varphi (z)$, as $z\downarrow 0$, is given by
  \begin{equation}
    \label{phiasymptn}
    \varphi (z)=\sum ^{n}_{k=0}\alpha _{k}z^{k}+\lambda \, 
    \alpha _{n+1}z^{n+1}+o(z^{n+1})\, .
  \end{equation}
  Actually, equality (\ref{phiasymptn}) follows directly from Lemma
  \ref{L:asymptser}. In more detail, Lemma \ref{L:asymptser} gives,
  for the case $n>0$,
  \begin{eqnarray}
    \varphi (z) & = & \sum ^{n+1}_{k=0}\varphi _{k}z^{k}
    \sum^{n+1-k}_{j=0}\left( k+\frac{3}{4}\right) _{j}z^{j}+o(z^{n+1})
    \nonumber \\
    & = & \sum ^{n+1}_{m=0}\sum ^{m}_{k=0}\varphi _{k}
    \left( k+\frac{3}{4}\right) _{m-k}z^{m}+o(z^{n+1})\, .
    \label{eqsforphik}
  \end{eqnarray}
  The coefficients $\varphi _{k}$, as given in (\ref{coeffphik}), have
  been chosen so that the asymptotics (\ref{phiasymptn}) is satisfied.
  This is to say that equalling (\ref{eqsforphik}) to
  (\ref{phiasymptn}) leads to a system of linear equations on the
  coefficients $\varphi_{k}$ whose unique solution is exactly
  (\ref{coeffphik}) as follows from the identity
  \begin{displaymath} 
    \sum ^{m}_{k=0}\alpha_{k}\, \left( k+\frac{3}{4}\right)_{m-k}
    -\sum ^{m}_{k=1}\alpha _{k-1}\, 
    \left( k-\frac{1}{4}\right) \, \left( k+\frac{3}{4}\right)_{m-k}
    =\alpha_{m}.
  \end{displaymath} 
  The case $n=0$ is even more straightforward.
  
  \emph{(iii)} $a_{+}\leq \alpha _{n+1}$ and $a_{-}\geq \alpha_{n+1}$.
  
  Let us show the first inequality, the other one can be proven
  analogously. We shall need the asymptotics of the function
  \begin{eqnarray*}
    \frac{(g(z)-1)^{2}}{g(z)\, z^{2}} & = & \left( 
      \sum ^{n}_{k=0}\alpha _{k}z^{k}+o(z^{n})\right)^{-1}
    \left( \sum ^{n}_{k=1}\alpha _{k}z^{k-1}+o(z^{n-1})\right)^{2}\\
    & = & \sum ^{n-1}_{k=0}\gamma _{k}z^{k}+o(z^{n-1})\, .
  \end{eqnarray*}
  Again, assumption (\ref{gasymptordn}) implies that
  $\sum^{n-1}_{k=0}\gamma _{k}z^{k}$ is a truncation of the power
  series
  \begin{eqnarray*}
    \frac{(\widetilde{g}(z)-1)^{2}}{\widetilde{g}(z)\, z^{2}} 
    & = & (\tilde{g}-1)\tilde{g}'+\frac{3}{4}
    \frac{\tilde{g}(\tilde{g}-1)}{z}\\
    & = & \frac{1}{2}(\tilde{g}^{2})'-\tilde{g}'+\frac{3}{4}\,
    \frac{\tilde{g}^{2}-\tilde{g}}{z}\, .
  \end{eqnarray*}
  Here we have used that $\tilde{g}$ solves (\ref{eqationB}).
  Combining (\ref{tildeg}) and (\ref{tildeg2}) one arrives at the
  formula
  \begin{equation}
    \label{coeffbetak}
    \gamma _{k}=\alpha _{k+2}-\left( k+\frac{7}{2}\right) \alpha _{k+1}.
  \end{equation}
  Now we can compare the functions $g(z)$ and $\varphi (z)$.  Let us
  choose $\lambda >1$ in (\ref{phiint0}) resp.  (\ref{phiintn}).
  Suppose that $\varphi (z)=g(z)$ at some point $z$.  Then, owing to
  (\ref{eqationB}) and (\ref{phidiffeq0}), it holds
  \begin{displaymath} 
    \varphi '(z)-g'(z)=\frac{1}{\lambda
      z^{2}g(z)}(g(z)-1)(g(z)-\lambda ),
  \end{displaymath} 
  when $n=0$, and using (\ref{phidiffeqn}), (\ref{coeffphik}) and
  (\ref{coeffbetak}),
  \begin{eqnarray*}
    \varphi '(z)-g'(z) & = & \frac{1}{z^{2}g(z)}(g(z)-1)^{2}
    -\sum ^{n+1}_{k=2}\varphi _{k}z^{k-2}\\
    & = & (1-\lambda )\alpha _{n+1}z^{n-1}+o(z^{n-1}),
  \end{eqnarray*}
  when $n>0$. In any case, there exists $\delta >0$ s.t.  $\varphi
  (z)=g(z)$ implies $\varphi '(z)<g'(z)$, $\forall z\in \, ]0,\delta
  [$ (in the case $n=0$, we need also that $g(z)>1$ if $z$ is
  sufficiently close to 0, see Proposition \ref{L:limg}). This means
  that the functions $g(z)$ and $\varphi(z)$ coincide in at most one
  point $z\in \, ]0,\delta [$.  Furthermore, we have already shown
  that $a_{-}\leq \alpha _{n+1}$, and so, using also
  (\ref{phiasymptn}), we conclude that
  \begin{displaymath} 
    \liminf_{z\downarrow 0}\frac{g(z)-\varphi (z)}{z^{n+1}}=a_{-}-\lambda\,
    a_{n+1}\leq (1-\lambda )\alpha _{n+1}<0.
  \end{displaymath} 
  Thus there exists a sequence $\{z_{n}\}$ s.t.  $z_{n}\downarrow 0$
  and $g(z_{n})<\varphi (z_{n})$, $\forall n$. Consequently,
  $g(z)<\varphi (z)$ on a right neighbourhood of 0. Hence, in virtue
  of (\ref{phiasymptn}),
  \begin{displaymath} 
    a_{+}\leq \limsup _{z\downarrow 0}\frac{\varphi(z)-\sum ^{n}_{k=0}
      \alpha _{k}z^{k}}{z^{k+1}}=\lambda \, \alpha_{n+1}.
  \end{displaymath} 
  The claim is a consequence of the limit $\lambda \downarrow 1$.
\end{proof}
\begin{cor}
  \label{C:smoothg}
  The left maximal solution $g(z)$, after having been defined at $z=0$
  by $g(0)=1$, belongs to $C^{\infty }([\, 0,z_{0}\,])$.
\end{cor}
\begin{proof}
  Observe that consecutive differentiation of equation
  (\ref{eqationB}) jointly with Proposition \ref{P:asymptofg} imply
  that, for any $m\in \Z _{+}$, $z^{m+1}g^{(m)}(z)$ has an asymptotic
  series at $z=0$ which we shall call $\sum ^{\infty }_{k=0}\alpha
  _{k}^{m}z^{k}\, .$We have to show that $g\in C^{m}$, $\forall m$,
  and this in turn amounts to showing that $\alpha ^{m}_{k}=0$ for
  $k<m+1$. Let us proceed by induction in $m$. The case $m=0$ was the
  content of Proposition \ref{L:limg}. Assume now that $g\in C^{m}$.
  Then $\alpha ^{m}_{k}=0$ for $k<m+1$, and the mean value theorem
  implies that
  \begin{equation}
    \label{meanvalthm}
    \liminf _{z\downarrow 0}g^{(m+1)}(z)\leq \frac{dg^{(m)}(0_{+})}{dz}
    =\alpha ^{m}_{m+2}\leq \limsup _{z\downarrow 0}g^{(m+1)}(z)\, .
  \end{equation}
  On the other hand, since $z^{m+2}g^{(m+1)}(z)$ has an asymptotic
  series the limit $\lim _{z\downarrow 0}g^{(m+1)}(z)$ always exists
  and equals either $\pm \infty$ or $\alpha ^{m+1}_{m+2}$ depending on
  whether there exists an index $k<m+2$ s.t.  $\alpha_{k}^{m+1}\neq 0$
  or not. However the property (\ref{meanvalthm}) clearly excludes the
  first possibility.
\end{proof}

\section{\label{sec:Asympt_of_h(t)}Asymptotics of a solution 
  \protect$h(t)\protect$ of the second order differential equation}

Except of the last subsection, we still consider the particular case
when $t_0=0$ and $h_1>0$ (see the remark at the end of 
Section~\ref{sec:Basic_properties}). We shall proceed to the case of
general initial condition only at the very end of the proof, in 
Subsection~\ref{sec:general-init-cond}.

\subsection{Reduction of the second order differential equation
  \label{sec: reduction_second_order_DE}}

Let us now complete some computations concerning the reduction of the
second order differential equation (\ref{eq:diff_eq_for_h}),
(\ref{eq:init_conds_for_h}) to a first order differential equation.
Let $g(z)$ be the left maximal solution of the first order
differential equation
\begin{displaymath} 
  \left( 1-\frac{3}{4}z\, g(z)-z^{2}g'(z)\right)
  g(z)=1,\textrm{ }g(z_{0})=g_{0},
\end{displaymath} 
where
\begin{displaymath}
  z_{0}=\frac{4}{h_{0}^{4}},\textrm{ }g_{0}=h_{0}^{3}h_{1}\, .
\end{displaymath} 
>From Section \ref{sec:A_reduced_DE_1st_order} we know that $g(z)$ is a
positive function from the class $C^{\infty }([\, 0,z_{0}\, ])$
(Proposition \ref{L:limg} and Corollary \ref{C:smoothg}). Consider the
function
\begin{equation}
  \label{defG}
  G(x)=\int ^{x}_{h_{0}^{4}}\frac{ds}{g\left( \frac{4}{s}\right) }\, ,
  \textrm{ }h_{0}^{4}\leq x<\infty .
\end{equation}
Then $G\in C^{\infty }([\,h_{0}^{4},\infty[\,)$, $G$ is strictly
increasing, $G(h_{0}^{4})=0$, and, owing to (\ref{limg}), $\lim_{x\to
  \infty }G(x)=\infty$. So the inverse function satisfies $G^{-1}\in
C^{\infty }([\, 0,\infty [)$ with $G^{-1}(0)=h_{0}^{4}$. Set
\begin{displaymath} 
  h(t)=\left(G^{-1}\left( 4t\right)\right)^{1/4}.
\end{displaymath} 
Then $h(t)$ solves the problem (\ref{eq:diff_eq_for_h}),
(\ref{eq:init_conds_for_h}).

Actually, $G(h(t)^{4})=4t$, $G'(h^{4})=g(4/h^{4})^{-1}$, and so
\begin{equation}
  \label{h3h'}
  h^{3}h'=\frac{1}{4G'(h^{4})}\frac{d\, G(h^{4})}{dt}
  =g\left( \frac{4}{h^{4}}\right) \, .
\end{equation}
Differentiating (\ref{h3h'}) once more gives
\begin{displaymath}
  3h^{2}(h')^{2}+h^{3}h''=-16\, h^{-5}h'g'\left( \frac{4}{h^{4}}\right) \, .
\end{displaymath} 
Denote for brevity $z=4/h^{4}$. Hence
\begin{eqnarray*}
  h^{3}(h'+h'') & = & g(z)-3h^{-4}(h^{3}h')^{2}-16\, h^{-8}(h^{3}h')g'(z)\\
  & = & g(z)-\frac{3}{4}z\, g(z)^{2}-z^{2}g(z)\, g'(z)\\
  & = & 1\, .
\end{eqnarray*}
Furthermore, $h(0)=\left(G^{-1}(0)\right)^{1/4}=h_{0}$ and
\begin{eqnarray*}
  h'(0) & = & G^{-1}(0)^{-3/4}\, \frac{d}{ds}G^{-1}(0)\, 
  =\, h_{0}^{-3}\, \left( \frac{d}{dx}G(h_{0}^{4})\right) ^{-1}\\
  & = & h_{0}^{-3}\, g\left( \frac{4}{h_{0}^{4}}\right) \, =\, h_{1}\, .
\end{eqnarray*}

\subsection{Asymptotics of \protect$G(x)\protect$}

First let us find, in $\C [[z]]$, the reciprocal element to the formal
power series $\tilde{g}(z)=\sum _{k=0}^{\infty }\alpha_{k}z^{k}$
defined in (\ref{tildeg}), (\ref{recurcoeff}). Set
\begin{displaymath}
  \tilde{g}(z)^{-1}=\sum ^{\infty }_{n=0}\beta _{n}z^{n}\, .
\end{displaymath} 
The formal power series $\tilde{g}(z)$ solves the differential
equation (\ref{eqationB}) and so an easy calculation shows that
$\tilde{f}(z)=\tilde{g}(z)^{-1}$ solves the differential equation
\begin{displaymath}
  f'(z)=-\frac{1}{z^{2}}\, f(z)^{2}(1-f(z))+\frac{3}{4}\frac{f(z)}{z}\,.
\end{displaymath} 
On the other hand, this differential equation implies a recursive rule
on the coefficients $\beta _{n}$, namely rule (\ref{eq:def-seq_beta})
preceding the formulation of Theorem \ref{thm:MainTheorem}.

\begin{lem}
  \label{L:asympt_of_G}
  The asymptotic series at infinity of the function $G(x)$ defined in
  (\ref{defG}) is given by
  \begin{displaymath} 
    G(x)\sim x-3\, \ln (x)+c-4\sum ^{\infty }_{k=1}
    \frac{\beta_{k+1}}{k} \left( \frac{4}{x}\right)^{k}
  \end{displaymath} 
  where
  \begin{equation}
    \label{eq:const_c}
    c =\int ^{\infty }_{h_{0}^{\, 4}}\left( \frac{1}{g
        \left( \frac{4}{s}\right) }-1+\frac{3}{s}\right) ds
    -h^{\, 4}_{0}+3\, \ln (h^{\, 4}_{0}).
  \end{equation}
\end{lem}
\begin{proof}
  It holds
  \begin{eqnarray*}
    G(x) & = & \int ^{\infty }_{h_{0}^{\, 4}}\left( \frac{1}{g
        \left( \frac{4}{s}\right) }-1+\frac{3}{s}\right) ds
    +\int ^{x}_{h_{0}^{\, 4}}\left( 1-\frac{3}{s}\right) ds\\
    & & -\int ^{\infty }_{x}\left(\frac{1}{g\left(\frac{4}{s}\right)} 
      -1+\frac{3}{s}\right) ds\, .
  \end{eqnarray*}
  According to Proposition \ref{P:asymptofg} we have the asymptotics
  at infinity,
  \begin{displaymath} 
    \frac{1}{g\left( \frac{4}{s}\right) }-1+\frac{3}{s}\sim
    \sum ^{\infty }_{k=2}\beta _{k}\left( \frac{4}{s}\right)^{k}\, .
  \end{displaymath}
  The claim then follows straightforwardly.
\end{proof}

\subsection{\label{sec:Asymptotics_Ginv}Asymptotics of 
  \protect$G^{-1}(x)\protect$}

Let us now focus on the inverse function $G^{-1}$.
\begin{lem}
  \label{L:estimGinv(x)}
  There exists $x_{\star }$ such that for all $x>x_{\star }$ it holds
  true that
  \begin{equation}
    \label{estim_on_Ginv(x)}
    0\leq G^{-1}(x)-x\leq \frac{x}{x-4}\, (x-G(x)).
  \end{equation}
\end{lem}
\begin{proof}
  Choose $y_{\star }\geq G(h_{0}^{\, 4})>0$ so that, for all $y\geq
  y_{\star }$,
  \begin{equation}
    \label{cond_on_ystar}
    0\leq 1-\frac{1}{g\left( \frac{4}{y}\right) }\leq 
    \frac{4}{y}\, \textrm{ and }\, y-G(y)\geq 0.
  \end{equation}
  This is possible owing to Proposition \ref{P:asymptofg} and Lemma
  \ref{L:asympt_of_G}. Set $x_{\star }=\max \{4,y_{\star }\}$.  For
  $x>x_{\star }$ fixed define a sequence $\{y_{n}\}_{n=0}^{\infty }$
  by the recursive rule
  \begin{displaymath} 
    y_{0}=x,\textrm{ }y_{n+1}=x+y_{n}-G(y_{n}).
  \end{displaymath} 
  Owing to our choice, $y_{n}\geq x\geq x_{\star }\geq y_{\star }$ for
  all $n$.  Using (\ref{defG}) one finds that
  \begin{displaymath} 
    y_{n+2}-y_{n+1}=y_{n+1}-y_{n}
    -\int^{y_{n+1}}_{y_{n}}\frac{ds}{g\left( \frac{4}{s}\right) }
    =\int^{y_{n+1}}_{y_{n}}
    \left( 1-\frac{1}{g\left( \frac{4}{s}\right)}\right) ds.
  \end{displaymath} 
  In virtue of (\ref{cond_on_ystar}), the sequence satisfies
  \begin{displaymath} 
    0\leq y_{n+2}-y_{n+1}\leq 4(\ln (y_{n+1})-\ln (y_{n}))\leq
    \frac{4}{x}\, (y_{n+1}-y_{n}).
  \end{displaymath} 
  Since $y_{1}-y_{0}=x-G(x)$ we get
  \begin{displaymath} 
    0\leq y_{n+1}-y_{n}\leq \left( \frac{4}{x}\right)^{n}(x-G(x)),
    \textrm{ }\forall n.
  \end{displaymath} 
  By the choice of $x_{\star }$ we have $4<x$ and, consequently, the
  sequence $\{y_{n}\}$ is convergent. The limit $y=\lim y_{n}$ solves
  $0=x-G(y)$ and so $y=G^{-1}(x)$. Moreover,
  \begin{eqnarray*}
    0\, \, \leq \, \, y-y_{n} & = & \sum ^{\infty }_{k=n}(y_{k+1}-y_{k})\\
    & \leq & \sum ^{\infty }_{k=n}\left(
      \frac{4}{x}\right)^{k}(x-G(x))
    =\frac{x}{x-4}\, (x-G(x))\left( \frac{4}{x}\right) ^{n}.
  \end{eqnarray*}
  The particular case $n=0$ in this relation is nothing but our claim.
\end{proof}
Combining Lemma \ref{L:estimGinv(x)} with Lemma \ref{L:asympt_of_G}
one immediately gets
\begin{cor}
  \label{C:asymptGinv-lowestorder}
  It holds true that, as $x\to +\infty $,
  \begin{displaymath}
    G^{-1}(x)=x+O(\ln (x)).
  \end{displaymath}
\end{cor}
Recall that in (\ref{eq:def-sigma0}), (\ref{eq:def-sigma_m}) we have
introduced polynomials $\sigma ^{m}_{k}(a_{1},a_{2},\ldots ,a_{k})$
labeled by indices $m\geq 0$ and $k\geq 1$.

\begin{prop}
  For all $n\in \Z _{+}$ it holds true that, as $x\to \infty$,
  \begin{equation}
    \label{asympt_of_Ginv-n}
    G^{-1}(x)=x+p_{0}(c;\ln (x))+\sum ^{n}_{k=1}
    \frac{p_{k}(c;\ln (x))}{x^{k}}
    +O\Big(\left( \frac{\ln (x)}{x}\right) ^{n+1}\Big )
  \end{equation}
  where the polynomials $p_{n}(c;z)$ have been defined in
  (\ref{eq:def-p0}), (\ref{eq:def-p_n}) and the constant $c$ is given
  by equality (\ref{eq:const_c}).
\end{prop}
\begin{proof}
  Corollary \ref{C:asymptGinv-lowestorder} implies
  \begin{equation}
    \label{asympt-lnGinv-reciGinv}
    \ln \left( G^{-1}(x)\right) =\ln (x)
    +O\left( \frac{\ln (x)}{x}\right) ,\textrm{ }\frac{1}{G^{-1}(x)}
    =\frac{1}{x}+O\left( \frac{\ln (x)}{x^{2}}\right) .
  \end{equation}
  Combining (\ref{asympt-lnGinv-reciGinv}) with Lemma
  \ref{L:asympt_of_G} one derives the relation
  \begin{equation}
    \label{x_eq_Ginv(x)_plusatd}
    x=G^{-1}(x)-3\, \ln (G^{-1}(x))+c
    -4\sum^{n}_{k=1}\frac{\beta _{k+1}}{k}\, 
    \left( \frac{4}{G^{-1}(x)}\right) ^{k}+O\left( \frac{1}{x^{n+1}}\right) ,
  \end{equation}
  valid for all $n\geq 0$. Setting $n=0$ in
  (\ref{x_eq_Ginv(x)_plusatd}) one arrives at the case $n=0$ in
  (\ref{asympt_of_Ginv-n}).  To finish the proof one can proceed, in
  the obvious way, by induction in $n$ when repeatedly using relation
  (\ref{x_eq_Ginv(x)_plusatd}).
\end{proof}

\subsection{Asymptotics of \protect$h(t)\protect$ for particular
  initial data}

We already know that $h(t)=\left(G^{-1}(4t)\right)^{1/4}$ solves the
problem $h(t)^{3}(h''(t)+h'(t))=1$, $h(0)=h_{0}$, $h'(0)=h_{1}$.
Using the known asymptotics of $G^{-1}(x)$ we get
\begin{displaymath}
  h(t)=(4t)^{1/4}\left( 1+\sum ^{n}_{k=1}
    \frac{p_{k-1}(c;\ln(4t))}{4^{k}t^{k}}
    +O\left( \frac{\ln (t)^{n}}{t^{n+1}}\right)\right)^{1/4}
\end{displaymath} 
and consequently
\begin{displaymath} 
  h(t)=(4t)^{1/4}\left( 1+\sum^{n}_{k=1}
    \frac{q_{k-1}(c;\ln (4t))}{t^{k}}
    +O\Big (\left( \frac{\ln(t)}{t}\right) ^{n+1}\Big )\right) 
\end{displaymath} 
where
\begin{displaymath} 
  q_{k}=\sum^{k}_{m=1}\frac{1}{4^{k}}\,
  \binom{\frac{1}{4}}{m}s_{m,k}(p_{0},p_{1},\ldots ,p_{k-1})\, .
\end{displaymath} 
So $q_{k}$ are exactly the polynomials introduced in Theorem
\ref{thm:MainTheorem}. It is also easy to see that the degree of
$q_{k}(c;z)$ is less than or equal to $k$ since the same is true for 
the polynomials $p_{n}$ with $n\geq 1$ and $\deg p_{0}=1$. This
observation in fact completes the proof of Theorem
\ref{thm:MainTheorem} in the case when $t_0=0$ and $h_1>0$.

\subsection{General initial conditions}\label{sec:general-init-cond}

Consider first a solution \( h(t) \) of (\ref{eq:diff_eq_for_h}) 
with the initial conditions \( h(0)=h_{0} \), \( h'(0)=h_{1} \), 
assuming that \( h_{1} \) is positive. Then, as we already know, 
the asymptotic behaviour of \( h(t) \) is described by 
Theorem~\ref{thm:MainTheorem}, i.e. equality (\ref{eq:asympt_h(t)}) 
holds true with \( c=c(0,h_{0},h_{1}) \). Choose
\( s\in \R  \) and set \( \tilde{h}(t)=h(t+s) \). Then \( \tilde{h}(t) \)
solves equation (\ref{eq:diff_eq_for_h}) 
and satisfies the initial conditions 
\( \tilde{h}(0)=\tilde{h}_{0}=h(s) \),
\( \tilde{h}'(0)=\tilde{h}_{1}=h'(s) \). But \( h'(s) \) is
positive for \( s \) sufficiently small and so 
equality (\ref{eq:diff_eq_for_h}) applies
to \( \tilde{h}(t) \) as well, with \( c \) being replaced
by \( \tilde{c}=c(0,\tilde{h}_{0},\tilde{h}_{1}) \). Equating
the asymptotics of \( h(t+s) \) to the asymptotics of \( \tilde{h}(t) \)
one arrives at the equality
\begin{equation}
  \label{eq:asymt_shiftbys}
  \begin{aligned}
   {} & \bigl (4(t+s)\bigr )^{1/4}\left( 1+\sum ^{n}_{k=1}
      \frac{q_{k}(c;\ln \bigl (4(t+s)\bigr ))}{(t+s)^{k}}\right) \\
   {} &\quad\qquad =(4t)^{1/4}\left( 1+\sum ^{n}_{k=1}
      \frac{q_{k}(\tilde{c};\ln (4t))}{t^{k}}
      +O\Big (\left( \frac{\ln (t)}{t}\right) ^{n+1}\Big )\right)
  \end{aligned}
\end{equation}
valid for \( t\to +\infty  \) and every \( n\in \Z _{+} \).
>From (\ref{eq:asymt_shiftbys}) it is not difficult to derive
the relation between \( c \) and \( \tilde{c} \), it reads
\begin{equation}
  \label{eq:ctilde}
  \tilde{c}=c-4s.
\end{equation}
Thus the invariance of the differential equation
(\ref{eq:diff_eq_for_h}) is reflected in an invariance of the 
asymptotic expansion of its solutions, as expressed by relations 
(\ref{eq:asymt_shiftbys}), (\ref{eq:ctilde}). 
It is also clear that these relations must hold true not only for 
\( s \) small but even for all \( s\in \R  \).

Choose now arbitrary initial data 
\( (t_{0},h_{0},h_{1})\in \R \times \, ]0,\infty [\, \times \R  \)
and let \( h(t) \) be the corresponding solution. Then, as we
know from Corollary \ref{cor:hprime_positive}, 
\( h'(t)>0 \) for all sufficiently
large \( t \). Fix \( s>t_{0} \) such that \( h'(s)>0 \) and
set \( \tilde{h}(t)=h(t+s) \). We use once more the already
proven fact that \( \tilde{h}(t) \) satisfies equality 
(\ref{eq:asympt_h(t)}), with
\( c \) being replaced by \( \tilde{c}=c(0,h(s),h'(s)) \).
This implies that the asymptotic behaviour of \( h(t)=\tilde{h}(t-s) \)
is given by
\[
h(t)=\bigl (4(t-s)\bigr )^{1/4}\left( 1+\sum ^{n}_{k=1}
  \frac{q_{k}(\tilde{c};\ln \bigl (4(t-s)\bigr ))}{(t-s)^{k}}
  +O\Big (\left( \frac{\ln (t)}{t}\right) ^{n+1}\Big )\right) ,
\]
\( n\in \Z _{+} \). But in that case one deduces from 
(\ref{eq:asymt_shiftbys}), (\ref{eq:ctilde}) that \( h(t) \) 
satisfies equality (\ref{eq:asympt_h(t)}) as well, with 
\( c=\tilde{c}+4s \). Theorem~\ref{thm:MainTheorem} is proven.

\section{\label{sec:Add_remark}Additional remark: comparison with the 
  asymptotics of 
  \emph{\protect$-W_{-1}(-e^{-x})\protect$}}

This is a digression whose aim is to emphasize a rather close similarity 
of the asymptotic behaviour of the function $G^{-1}(x)$ 
with that of Lambert function.
The Lambert function $W(z)$ gives the principal solution for $w$ in
$z=we^{w}$ and $W_k(z)$ gives the $k^{\mathrm{th}}$ solution.
Surprisingly, it is not documented in some standard 
text books and reference books on special functions
though we may have missed some sources. On the other hand, 
the Lambert function seems to have attracted even in a rather recent 
period some attention, particularly from the computational and 
combinatorial point of view (see \cite{LambertW} for a summary). It is 
also implemented in some standard computer algebra systems like Maple
and Mathematica where it is called LambertW and ProductLog, 
respectively. Let us just briefly recall that 
$W(z)$ is analytic in a neighbourhood of $z=0$ with
the convergence radius equal to $e^{-1}$,
\begin{displaymath} 
  W(z)=\sum ^{\infty }_{k=1}\frac{(-1)^{k}k^{k-1}}{k!}\, z^{k}\, .
\end{displaymath} 
The coefficients have a combinatorial interpretation when counting
distinct oriented trees.

Consider now the equation
\begin{displaymath} 
  y-\ln (y)=x,
\end{displaymath} 
or, equivalently,
\begin{displaymath}
  ye^{-y}=e^{-x}\, .
\end{displaymath} 
It is elementary to see that for $x\in \, ]1,+\infty [$ there are
exactly two real solutions, $y_{1}(x)$ and $y_{2}(x)$, with
$y_{1}(x)\in \, ]0,1[\,$ and $y_{2}(x)\in \, ]1,+\infty [\,$.  The
both solutions can be expressed with the aid of the Lambert function, 
namely
\begin{displaymath} 
  y_{1}(x)=-W(-e^{-x}),\quad y_{2}(x)=-W_{-1}(-e^{-x})\, .
\end{displaymath} 

The aim of this remark is to point out that the asymptotics of the 
second solution, i.e. $-W_{-1}(-e^{-x})$, as $x\to +\infty$,
can be derived in a way quite similar to what we have done 
in Subsection \ref{sec:Asymptotics_Ginv} when treating the function 
$G^{-1}(x)$.  To this end let us recursively define polynomials 
$\tilde{p}_{k}(z)$,
\begin{displaymath}
  \tilde{p}_{0}(z)=z,\ \tilde{p}_{k+1}(z)
  =\sigma^{0}_{k+1}(\tilde{p}_{0}(z),\tilde{p}_{1}(z),\dots,
  \tilde{p}_{k}(z)).
\end{displaymath} 
For $k\geq 1$, the degree of the polynomial $\tilde{p}_{k}(z)$ is $k$.
Here are several first polynomials:
\begin{displaymath} 
  \tilde{p}_{1}(z)=z,\textrm{  }\tilde{p}_{2}(z)
  =z-\frac{1}{2}\, z^{2},\textrm{  }\tilde{p}_{3}(z)
  =z-\frac{3}{2}\, z^2+\frac{1}{3}\, z^{3},\textrm{  }\ldots .
\end{displaymath}

\begin{prop}
  It holds, as $x\to +\infty$,
  \begin{displaymath} 
    -W_{-1}(-e^{-x})=x+O(\ln x)
  \end{displaymath}
  and, for $n\geq 0$,
  \begin{displaymath} 
    -W_{-1}(-e^{-x})=x
    +\sum^{n}_{k=0}\frac{\tilde{p}_{k}(\ln x)}{x^{k}}
    +O\left( \left( \frac{\ln x}{x}\right) ^{n+1}\right) .
  \end{displaymath}
\end{prop}
The proposition can be proven using a similar approach as the one used
in the proof of Lemma \ref{L:estimGinv(x)}. In fact, this asymptotic 
expansion is well known and is in agreement with what has been published 
in \cite{Bruijn, Comtet} and \cite{LambertW} though the derivation 
and presentation here is somewhat different.

\vskip 24pt
\noindent{\bf Acknowledgements.}
R.D.B. wishes to thank FONDECYT (Chile) 199--0427, and the action 
C94E10 of ECOS-CONICYT.
P.\v{S}. wishes to gratefully acknowledge the partial support from 
Grant No. 201/01/01308 of Grant Agency of the Czech Republic.


\begin{thebibliography}{1}
\bibitem{Abraham-Marsden}Abraham R., Marsden J.E.: \emph{Foundations
    of Mechanics}. 
  \newblock Addison-Wesley, 1978.  
\bibitem{Halperin} Halperin B.:
  \newblock Phys. Rev. B {\bf 25} (1982)  2185ff.
\bibitem{LambertW} Corless R.M., Gonnet G.H., Hare D.E.G., 
Jeffrey D.J., Knuth D.E.: \emph{On the Lambert W Function}. 
  \newblock Advances in Computational Mathematics {\bf 5} (1996) 329-359.
\bibitem{Bruijn} de Bruijn N.G.: \emph{Asymptotic Methods in Analysis}. 
  \newblock North-Holland, 1961.
\bibitem{Comtet} Comtet L.: 
  \newblock C. R. Acad. Sc. Paris {\bf 270} (1970) 1085-1088.

\end{thebibliography}
\end{document}